\documentclass[a4paper,12pt]{article}
\usepackage{graphicx,amssymb,bm,latexsym,color,mathrsfs,textcomp}
\usepackage{amsmath}
\makeatletter
\@addtoreset{equation}{section}

\makeatother
\newdimen\argwidth
\def\db[#1\db]{%
 \setbox0=\hbox{$\displaystyle #1$}\argwidth=\wd0
 \setbox0=\hbox{$\left[\box0\right]$}\advance\argwidth by -\wd0
 \left[\kern.3\argwidth\box0 \kern.3\argwidth\right]}
\def\sqr#1#2{{\vcenter{\hrule height.#2pt
      \hbox{\vrule width.#2pt height#1pt \kern#1pt
          \vrule width.#2pt}
      \hrule height.#2pt}}}
\renewcommand{\Box}{\,{\mathchoice{\sqr84}{\sqr84}{\sqr{5.0}3}{\sqr{3.5}3}}\,}
\newcommand{\bea}{\begin{eqnarray}}
\newcommand{\ena}{\end{eqnarray}}
\newcommand{\vs}[1]{\vspace{#1 mm}}
\newcommand{\hs}[1]{\hspace{#1 mm}}
\renewcommand{\a}{\alpha}
\renewcommand{\b}{\beta}
\renewcommand{\c}{\gamma}

\renewcommand{\d}{\delta}
\newcommand{\e}{\epsilon}
\newcommand{\s}{\sigma}

\newcommand{\la}{\lambda}
\newcommand{\pa}{\partial}

\newcommand{\nn}{\nonumber\\}
\newcommand{\p}[1]{(\ref{#1})}

\newcommand{\bg}{\bar g}

\def\calA{{\cal A}}
\def\calB{{\cal B}}
\def\calC{{\cal C}}
\def\calD{{\cal D}}
\def\calE{{\cal E}}
\def\calP{{\cal P}}
\def\calQ{{\cal Q}}
\def\II{I\!I}

\def\B{{\rm B}}
\def\disp{\displaystyle}
\def\tr{\mathop{\hbox{tr}}}

\begin{document}

\begin{titlepage}

\begin{flushright}
MISC-2014-01\\
KU-TP 061 \\
\today
\end{flushright}

\vs{10}
\begin{center}
{\Large\bf Covariant Approach to the No-ghost Theorem \\
in Massive Gravity}
\vs{15}

{\large
Taichiro Kugo\footnote{e-mail address: kugo@cc.kyoto-su.ac.jp}$^{,a}$
and
Nobuyoshi Ohta\footnote{e-mail address: ohtan@phys.kindai.ac.jp}$^{,b}$
} \\
\vs{10}

$^a${\em Department of Physics and
Maskawa Institute for Science and Culture, \\
Kyoto Sangyo University, Kyoto 603-8555, Japan}
\vs{2}

$^b${\em Department of Physics, Kinki University,
Higashi-Osaka, Osaka 577-8502, Japan}

\vs{15}
{\bf Abstract}
\end{center}

We discuss the no-ghost theorem in the massive gravity in a covariant manner.
Using the BRST formalism and St\"{u}ckelberg fields, we first clarify
how the Boulware-Deser ghost decouples in the massive gravity theory with Fierz-Pauli mass term.
Here we find that the crucial point in the proof is that there is no higher (time) derivative for
the St\"{u}ckelberg `scalar' field. We then analyze the nonlinear massive gravity proposed
by de Rham, Gabadadze and Tolley, and show that there is no ghost for general
admissible backgrounds.
In this process, we find a very nontrivial decoupling limit for general backgrounds.
We end the paper by demonstrating the general results explicitly in a
nontrivial example where there apparently appear higher time derivatives for
St\"{u}ckelberg scalar field, but show that this does not introduce the ghost
into the theory.

\end{titlepage}
\newpage
\setcounter{page}{2}

\section{Introduction}

Recently there has been renewed interest in the search for the modification
of gravity at large distances by adding the mass terms for graviton.
Motivation comes from both theoretical and observational considerations.

On the theoretical side, it is interesting to explore the possibility of formulating
theory of massive spin-2 field.
In general relativity which describes massless spin-2 field, the four constraints of
the theory together with the invariance under the four general coordinate transformations
remove eight of the modes from the ten degrees of freedom in the metric, and
the number of the propagating modes reduces to the physical two modes of massless
graviton. When the mass term is added, four constraints remove four propagating modes,
but the general covariance is broken. Thus there remain six degrees of freedom in general.
Five out of these are the modes of massive spin-2 graviton, but it turns out that
the sixth scalar mode is a ghost with a negative metric.
A unique mass term that does not contain this ghost has been known as Fierz-Pauli mass
term~\cite{FP}.

However it has been pointed out that this theory suffers from the problem
that the helicity-0 state couples to the trace of the matter stress-energy tensor with
the same strength as the helicity-2 state~\cite{disc}. This means that this
massive gravity does not continuously reduce to general relativity in the massless limit.
This is called vDVZ discontinuity.
It was then argued by Vainshtein that the discontinuity could be avoided by the nonlinear
interaction~\cite{V}.
Unfortunately the very nonlinearity that cures the discontinuity problem
re-introduces the ghost in the theory, named Boulware-Deser (BD) ghost~\cite{BD}.
It is a major theoretical problem how to extend the mass term consistently
to nonlinear level.

On the observational side, the recent discovery of the accelerating expansion of
the universe suggests the modification of either the gravity side or matter side
of the Einstein equation. A simple extension would be to introduce the cosmological constant,
which must be extremely tiny to account for the current observation.
Another modification on the gravity side is to consider the massive gravity,
because cosmological solutions with an accelerated expansion are expected
if the gravity becomes weaker on the larger scale.

Recently an interesting proposal to extend the work of Fierz-Pauli~\cite{FP}
to the nonlinear level has been made by
de Rham, Gabadadze and Tolley (dRGT)~\cite{DG,DGT1}
by generalizing the effective
field theory approach~\cite{AGS}. It was first shown that there is no BD ghost
to all orders in the decoupling limit (defined in the flat space).
It has then been argued that this formulation of massive gravity has no
ghost at nonlinear level~\cite{HR1}--\cite{DMT1}.
Using the noncovariant Arnowitt-Deser-Misner (ADM) decomposition, it is shown that the mass term
introduces nonlinear terms for the shifts so that these do not produce any
constraint, but the lapse function remains linear and we are left with one constraint
instead of four in general relativity. Thus, in this noncovariant approach, we have six degrees of
freedom for the propagating modes from the spatial metric $g_{ij}$, but one of them
is removed by the above constraint from the lapse, leaving correct five degrees of freedom
for a massive spin-2 without ghost. The proof is valid to full nonlinear level,
but it is based on noncovariant formulation and is very indirect one
just counting degrees of freedom. So the reason is left unclear
why there remains such a linear lapse variable in the dRGT massive gravity.

An interesting approach is the one to introduce St\"{u}ckelberg fields which recover
the general coordinate invariance~\cite{AGS,DGT2,HR4} and additional gauge invariance.
Here again using ADM decomposition, it is shown that we get the right five physical
degrees of freedom in the theory.
This is again a noncovariant approach.

There is another covariant approach to the problem in~\cite{DMZ}, which uses again
constraints to remove
the degrees of freedom, but the proof is not completed for general mass terms,
in particular in the presence of cubic mass term.

In this paper we use the covariant approach based on the BRST formalism and
St\"{u}ckelberg fields to show explicitly the cancellation of the ghost
degrees of freedom 
and for arbitrary backgrounds, and clarify the structure of the theory.
We show that we have 10 degrees of freedom from the graviton, 4 from the St\"{u}ckelberg
vector and 1 from the St\"{u}ckelberg scalar field, minus $4\times2$ from the vector
Faddeev-Popov ghost and anti-ghost, minus $1\times2$ from the scalar Faddeev-Popov
ghost and anti-ghost.
This leaves us with 5 degrees of freedom, the right number for massive spin-2 fields.
An important point is that there is no higher derivative term for the kinetic
term of the St\"{u}ckelberg scalar field, which (if present) introduces additional ghost degree of
freedom unless the mass term is judiciously chosen.
It was shown for several cases that there is no such higher order term
or it is present but in a harmless way in the ADM formulation~\cite{DGT2}.
However it was not clear if this is true in general and for arbitrary backgrounds.
Here we give the complete proof of the absence of ghost with all possible
mass terms and on general backgrounds in a covariant manner.

This paper is organized as follows.
In sect.~\ref{fp:noghost}, in order to get the idea how our approach works, we show in detail how the BD
ghost is decoupled in the simple theory with Fierz-Pauli mass term in our formulation.
Since it is easy to do this for arbitrary dimensions, we discuss the problem in general
dimensions $D$.
First in sect.~\ref{fp:brs}, we discuss the streamlined proof of
the no-ghost theorem using the St\"{u}ckelberg fields and BRST formalism in this theory.
Here we make the counting of degrees of freedom, and clarify that the necessary and sufficient
condition for the theory to be ghost free is that there is no higher (time) derivative
of the St\"{u}ckelberg fields.
The discussion is completed in sect.~\ref{fp:propagator}, where we compute the propagators
for all the fields in the theory and show that all the ghost degrees of freedom cancel
against Faddeev-Popov ghosts, and there remain only physical $\frac{(D-2)(D+1)}{2}$
(five for four dimensions) degrees of freedom for spin-2.

In sect.~\ref{full}, we come to the main theme of this paper to prove the no-ghost theorem
in the nonlinear massive gravity on arbitrary backgrounds.
In sect.~\ref{dia}, we first discuss how to diagonalize general background metric
in order to properly take its square root which is necessary to write the mass term
suitable for examining the spectrum.
We then use this result in sect.~\ref{mass} to compute the generating function of the mass terms.
In sect.~\ref{hidden}, we find that there is an important hidden $U(1)$ gauge invariance which
ensures the decoupling of the ghost. In this process, we find that the way of how to introduce
the St\"{u}ckelberg fields in general background is significantly modified from the counterpart
for flat background, and the associated decoupling limit is also quite nontrivial.
We then show that there is no higher derivative terms for the St\"{u}ckelberg fields.
Combined with the above result in the Fierz-Pauli mass term, this implies that
there remains no BD ghost in this massive gravity.
In sect.~4, we go on to discuss an explicit and nontrivial example for a background
metric with shift. We show that naively it looks that there appears higher time derivatives
on the St\"{u}ckelberg scalar field, but our definition of the St\"{u}ckelberg fields avoids
the trouble, so that there is no BD ghost in the theory.

\section{Absence of ghost in Fierz-Pauli mass term}
\label{fp:noghost}

In this section, we first discuss the no ghost theorem in massive gravity with Fierz-Pauli
mass term in arbitrary dimensions $D$. Let us consider the action
\bea
S=\frac{1}{\kappa^2}\int d^D x \sqrt{- g} \Big[ R -\frac{m^2}{4} (h_{\mu\nu}^2 -a h^2 )\Big] ,
\label{action}
\ena
where $\kappa^2$ is the $D$-dimensional gravitational constant, $m$ and $a$ are constants.
Here $h_{\mu\nu}$ is the fluctuation of the metric around the background spacetime
\bea
g_{\mu\nu}= \bar g_{\mu\nu} + \kappa h_{\mu\nu},
\label{fluc}
\ena
and $h \equiv\bar g^{\mu\nu} h_{\mu\nu}$.
We use the conventions in Ref.~\cite{Ohta} and set $\kappa=1$ henceforth.
In the rest of this section, we consider the flat background
$\bar g_{\mu\nu}=\eta_{\mu\nu}$ for simplicity.

At first sight, one expects that this theory contains $\frac{(D-2)(D+1)}{2}$
(five for four dimensions) degrees of freedom corresponding to the massive spin-2 field.
However it has been known that this massive gravity contains an additional (sixth for
four dimensions) degree of freedom unless $a=1$, known as BD ghost~\cite{BD}.
We first recapitulate how to understand this situation.

\subsection{St\"{u}ckelberg fields and BRST formalism}
\label{fp:brs}

Because of the presence of mass term in \p{action}, there is no invariance under
the general coordinate transformation. We can recover the invariance
by introducing
the St\"{u}ckelberg fields as was shown by Arkani-Hamed, Georgi and
Schwartz~\cite{AGS}. In their formulation, $h_{\mu\nu}=g_{\mu\nu}-\eta_{\mu\nu}$ is
replaced by $h_{\mu\nu}=g_{\mu\nu}-f_{\mu\nu}$ where
$f_{\mu\nu}$ is the fiducial metric given as the general
coordinate transformation of the flat metric $\eta_{\mu\nu}$
using St\"{u}ckelberg fields (See the precise definition given later in sect.~3.)
This replacement reduces at the linearized level simply to
\cite{Hamamoto:2005in}
\bea
h_{\mu\nu} \Rightarrow h_{\mu\nu} -\frac{1}{m}(\pa_\mu A_\nu+\pa_\nu A_\mu)
+\frac{2}{m^2}\pa_\mu\pa_\nu\pi,
\label{stu}
\ena
where $m$ is a mass scale. The metric is invariant under the transformation
\bea
\d h_{\mu\nu} = \pa_\mu\xi_\nu+ \pa_\nu\xi_\mu, ~~~
\d A_\mu= m \xi_\mu+ \pa_\mu\Lambda, ~~~
\d\pi= m\Lambda,
\label{gt}
\ena
In order to quantize the theory, we gauge fix the theory and introduce the Faddeev-Popov
ghosts and anti-ghosts corresponding to the invariance~\p{gt}.
They are vector ghost
$c_\mu$ and anti-ghost $\bar c_\mu$ for $\xi_\mu$, and scalar ghost $c$ and
anti-ghost $\bar c$ for $\Lambda$.

Now the physical degrees of freedom in the theory is counted as
$\frac{D(D+1)}{2}$ (10 for four dimensions) from $h_{\mu\nu}$,
$D$ (4 for four dimensions) from $A_\mu$
and 1 from $\pi$, minus $D\times2$ ($4\times2$ for four dimensions) from the vector
ghost and anti-ghost,
minus 2 from the scalar ghost and anti-ghost.
This leaves us $\frac{(D-2)(D+1)}{2}$ (5 for four dimensions)
degrees of freedom, right number for massive spin-2 fields. However,
this cannot be true in general.
It has been known that the theory~\p{action} describes
$\frac{D(D-1)}{2}$ (6 for four dimensions)
degrees of freedom unless $a=1$ and one of them is a ghost. What is wrong with this counting then?

We can see the origin of the problem if we substitute \p{stu} into the action:
The quadratic part of the mass terms of the Lagrangian in~\p{action} takes the form
\bea
&&\hs{-5} -\frac{m^2}{4}( h_{\mu\nu}^2-a h^2) - \frac14(\pa_\mu A_\nu-\pa_\nu A_\mu)^2
-(1-a)(\pa_\mu A^\mu)^2 - (m A_\mu-\pa_\mu\pi)(\pa_\nu h^{\mu\nu}-\pa^\mu h) \nn
&& \hs{-5} -m(a-1) h \pa_\mu A^\mu+(a-1) h \Box \pi
+\frac{2(1-a)}{m}\pa_\mu A^\mu\Box\pi- \frac{(1-a)}{m^2} (\Box \pi)^2.
\ena
The last term here indicates that the field $\pi$ has two degrees of freedom
unless $a=1$, so the above counting is not correct. If and only if $a=1$,
the above counting
is correct and we are left with $\frac{(D-2)(D+1)}{2}$ (5 for four dimensions)
degrees of freedom.
Note that the terms involving only $\pi$ vanishes in this case, and this corresponds
to the requirement of no-ghost in the decoupling limit.
The mixing with the metric fluctuation gives the dynamics to $\pi$.
This can be checked by taking the determinant of kinetic matrix (containing only second
derivatives) to see if it does not vanish identically.

Alternatively we can see that the shift
$h_{\mu\nu} \to h_{\mu\nu}+\frac{2}{D-2} \eta_{\mu\nu} \pi$
in the Einstein term
\bea
{\cal L}_{E,2} = \frac14 h^{\mu\nu}\Big[ \pa_\mu\pa_\nu h-\pa_\mu h_\nu
-\pa_\nu h_\mu+\Box h_{\mu\nu} + \eta_{\mu\nu}(\pa_\la h^\la-\Box h)\Big],
\ena
cancels the mixing and produces normal kinetic term for $\pi$.
Here we have defined
\bea
h_\mu= \pa^\nu h_{\mu\nu},~~
h = h_\mu^\mu.
\ena
At this stage, we have
\begin{eqnarray}
{\cal L}_{E,2}+ {\cal L}_{\rm mass}
&\rightarrow&
{\cal L}_{E,2}
-{m^2\over4}(h_{\mu\nu}^2-h^2) +{D-1\over D-2}m^2 h\,\pi
+{D(D-1)\over(D-2)^2}m^2\pi^2 \nonumber\\
&& -{1\over4}F_{\mu\nu}(A)^2 -mA^\mu\left(h_\mu-\partial_\mu h
 -2{D-1\over D-2}\partial_\mu\pi\right)
+{D-1\over D-2} \pi\square\pi.~~~~~
\label{eq:2.18}
\end{eqnarray}
Here and henceforth in this section,
all the $h_{\mu\nu}$ and $h$ denote the new gravity
fields after the above shifting:
\begin{equation}
h_{\mu\nu} = h_{\mu\nu}^{\rm original} - \frac{2}{D-2} \eta_{\mu\nu} \pi,
\qquad
h\equiv\eta^{\mu\nu}h_{\mu\nu} = h^{\rm original} - \frac{2D}{D-2} \pi,
\end{equation}

We now discuss the gauge fixing of the theory and examine what spectrum
we have explicitly. In order to resolve the field mixing terms, we
adopt the so-called $R_\xi$ gauges.
The gauge fixing and Faddeev-Popov terms are concisely written as
\begin{eqnarray}
{\cal L}_{\rm GF}+{\cal L}_{\rm FP}
&=& -i \delta_{\B} \left[
\bar c^\mu\Big( h_\mu-x \partial_\mu h-\alpha m A_\mu+\frac{\alpha}{2}B_\mu\Big)
\right] \nn
&&{}-i \delta_{\B} \left[
\bar c \Big\{\partial A-m\beta(yh+z\pi)+\frac{\beta}{2}B \Big\} \right]\ ,
\label{gffp}
\ena
where $\a, \b, x,y,z$ are gauge parameters, and the (fermionic) BRST
transformations are defined as
\begin{eqnarray}
&& \delta_\B h_{\mu\nu} = \pa_\mu c_\nu+\pa_\nu c_\mu
 -{2\over D-2}\eta_{\mu\nu}mc ,~
\delta_\B A_\mu=  m c_\mu+ \pa_\mu c ,~
\delta_\B \pi= \, m c, \nn
&& \d_B c_\mu= c^\rho\pa_\rho c_\mu, ~
\d_B \bar c_\mu= i B_\mu, ~
\d_B B_\mu= 0,~
\d_B c =  c^\rho\pa_\rho c, ~
\d_B \bar c = i B, ~
\d_B B = 0.~~~
\end{eqnarray}
Using \p{gffp}, we find the gauge fixing and Faddeev-Popov terms as follows:
\bea
{\cal L}_{\rm GF} \hs{-2} &=& \hs{-2}
B^\mu\left(h_\mu-x \partial_\mu h-\alpha m A_\mu\right) +\frac{\alpha}2 B_\mu^2
+B\left(\partial A-m\beta(yh+z\pi)\right)+\frac{\beta}2 B^2 \nn
\hs{-2} &=& \hs{-2} -\frac1{2\a}[h_\mu-x \pa_\mu h-\alpha m A_\mu]^2
-\frac{1}{2\b} [\pa_\mu A^\mu-m\b(yh+z\pi)]^2+\frac{\alpha}2 B^{\prime 2}_\mu
+\frac{\b}2 B^{\prime 2},
\label{eq:GF} \\
{\cal L}_{\rm FP}
\hs{-2} &=& \hs{-2} i\bar c^\mu\left[ \partial_\mu\partial^\nu c_\nu
 + \square c_\mu-m{2\over D-2}\pa_\mu c
-x \partial_\mu(2\partial^\nu c_\nu- {2D\over D-2}mc) -\a m^2c_\mu-\a m\pa_\mu c\right] \nn
&& {}+i\bar c \left[m\pa^\nu c_\nu+\square c -m\b\left(y(2\pa^\nu c_\nu-{2D\over D-2}mc)
+zmc\right)\right].
\end{eqnarray}
Here $B'_\mu$ and $B'$ are the shifted $B_\mu$ and $B$ fields to complete the squares:
\begin{eqnarray}
B'_\mu&=&
B_\mu+ \alpha^{-1}\left(h_\mu-x \partial_\mu h-\alpha m A_\mu\right), \nn
B'&=&
B + \beta^{-1}\left(\partial A-m\beta(yh+z\pi)\right).
\end{eqnarray}
Now we determine the gauge parameters $x, y$ and $z$ so as to cancel the
various field transition terms as follows.
Note that this gauge fixing term (\ref{eq:GF}) is arranged to cancel
the term $-mA_\mu h^\mu$ in (\ref{eq:2.18}). In order to cancel the
$A_\mu\partial^\mu h$ term in (\ref{eq:2.18}) by the corresponding terms from
the gauge fixing term (\ref{eq:GF}), we should have
\begin{equation}
x+y=1 .
\end{equation}
To cancel $A_\mu\partial^\mu\pi$ term in (\ref{eq:2.18}) by a term from
(\ref{eq:GF}), we set
\begin{equation}
z= 2{D-1\over D-2}.
\end{equation}
Finally the $h\pi$ mixing term  in (\ref{eq:2.18}) can be cancelled by
that from (\ref{eq:GF}) by choosing
\begin{equation}
y = {1\over2\beta}.
\end{equation}
The resulting total quadratic Lagrangian is
\bea
{\cal L}_{t}\! &=&\! {\cal L}_{E,2}+ {\cal L}_{\rm mass}+{\cal L}_{\rm GF} \nn
\! &=&\! \Big(\frac12-\frac{x}{\alpha}\Big) h \partial_\mu h^\mu
+ \Big(\frac12-\frac{1}{2\a}\Big) h_\mu^2
+\frac14 h_{\mu\nu} (\Box -m^2) h^{\mu\nu}
- \Big(\frac14-\frac{x^2}{2\a}\Big) h\Box h \nn
&& \hs{-2}+\frac{xm^2}{4}  h^2
-\frac14 (\pa_\mu A_\nu-\pa_\nu A_\mu)^2
-\frac{1}{2\b}(\partial A)^2 -\frac12 \alpha m^2 A_\mu^2
+\frac{z}{2} \pi\Big[ \Box -2\beta w m^2 \Big]\pi,~~~~~
\ena
with the parameter $w$ denoting
\begin{equation}
w={xD-1\over D-2}.
\end{equation}
The action takes a simple form for $\a=\b=1$, in which case $x=1/2$ and
all the fields
have the same mass $m^2$.

\subsection{Propagators}
\label{fp:propagator}

\subsubsection{tensor propagator}

To calculate the propagators, let us introduce projection operators in
momentum space:
\begin{eqnarray}
d_{\mu\nu} &\equiv& \eta_{\mu\nu}-{p_\mu p_\nu\over p^2}, \qquad
e_{\mu\nu}\equiv{p_\mu p_\nu\over p^2}, \nn
I_{\mu\nu, \rho\sigma} &\equiv& \frac12 \Big( d_{\mu\rho}d_{\nu\sigma}
 +d_{\mu\sigma}d_{\nu\rho}-{2\over D-1}d_{\mu\nu}d_{\nu\sigma} \Big), \nn
\II_{\mu\nu, \rho\sigma} &\equiv& \frac12 [ d_{\mu\rho}e_{\nu\sigma} +d_{\mu\sigma}e_{\nu\rho}
+ (\mu\leftrightarrow \nu)],
\end{eqnarray}
which satisfy
\begin{eqnarray}
&&d_{\mu, \rho}\,d^{\rho}_{\ \ \nu}=d_{\mu\nu}, \quad
d_{\mu, \rho}\,e^{\rho}_{\ \ \nu}=
e_{\mu, \rho}\,d^{\rho}_{\ \ \nu}=0, \quad
e_{\mu, \rho}\,e^{\rho}_{\ \ \nu}=e_{\mu\nu}, \nn
&&I_{\mu\nu, \alpha\beta}\,I^{\alpha\beta}_{\ \ \ \rho\sigma}=I_{\mu\nu, \rho\sigma},
\qquad \II_{\mu\nu, \alpha\beta}\,\II^{\alpha\beta}_{\ \ \ \rho\sigma}
=\II_{\mu\nu, \rho\sigma}, \nn
&&I_{\mu\nu, \alpha\beta}\,\II^{\alpha\beta}_{\ \ \ \rho\sigma}=
\II_{\mu\nu, \alpha\beta}\,I^{\alpha\beta}_{\ \ \ \rho\sigma}=0, \nn
&&I_{\mu\nu, \rho\sigma}\,d^{\rho\sigma}=I_{\mu\nu, \rho\sigma}\,e^{\rho\sigma}=0, \qquad
\II_{\mu\nu, \rho\sigma}\,d^{\rho\sigma}=\II_{\mu\nu, \rho\sigma}\,e^{\rho\sigma}=0.
\end{eqnarray}
Note that
\begin{eqnarray}
h\partial_\mu h^\mu
&=& -\frac12 h^{\mu\nu}\, p^2 \left(d_{\mu\nu}e_{\rho\sigma} +e_{\mu\nu}d_{\rho\sigma}
+2e_{\mu\nu}e_{\rho\sigma} \right) h^{\rho\sigma}, \nn
h_\mu h^\mu
&=& +\frac12 h^{\mu\nu}\, p^2 \left( \II_{\mu\nu, \rho\sigma}
+ 2e_{\mu\nu}e_{\rho\sigma} \right) h^{\rho\sigma}, \nn
h_{\mu\nu}\square h^{\mu\nu}
&=& - h^{\mu\nu}\, p^2 \left( I_{\mu\nu, \rho\sigma}+\II_{\mu\nu, \rho\sigma}
+{1\over D-1}d_{\mu\nu}d_{\rho\sigma} +e_{\mu\nu}e_{\rho\sigma} \right) h^{\rho\sigma},\nn
h\square h
&=& - h^{\mu\nu}\, p^2 \Bigl(
d_{\mu\nu}d_{\rho\sigma} + (e_{\mu\nu}d_{\rho\sigma}
+d_{\mu\nu}e_{\rho\sigma}) +e_{\mu\nu}e_{\rho\sigma} \Bigr) h^{\rho\sigma}.
\end{eqnarray}
Using these, we find that the quadratic term in the gravity field
$h_{\mu\nu}$ is written in the form
\begin{eqnarray}
\frac12 h^{\mu\nu}\, \calQ_{\mu\nu, \rho\sigma}\, h^{\rho\sigma},
\ena
where
\bea
\calQ_{\mu\nu, \rho\sigma} =
\calA\,I_{\mu\nu, \rho\sigma}+\calB\,\II_{\mu\nu, \rho\sigma}
+\calC\,d_{\mu\nu}d_{\rho\sigma} + \calD\,(e_{\mu\nu}d_{\rho\sigma}
+d_{\mu\nu}e_{\rho\sigma}) +\calE\,e_{\mu\nu}e_{\rho\sigma},
\end{eqnarray}
with
\begin{eqnarray}
&& 2\calA= -(p^2+m^2),\quad
2\calB= -{p^2+\alpha m^2\over\alpha}, \\
&&\calC= {D-2\over2(D-1)}(p^2+2\beta wm^2) - {x^2\over\alpha}(p^2+\alpha\beta m^2),\quad
 \nn
&&\calD= {x\over2\alpha\beta}(p^2+\alpha\beta m^2), \quad
\calE= -{1\over4\alpha\beta^2}(p^2+\alpha\beta m^2), \nn
&&\calD^2-\calE\calC= {D-2\over8\alpha\beta^2(D-1)}(p^2+\alpha\beta m^2)(p^2+2\beta wm^2).
\end{eqnarray}

The propagator $\calP$
\begin{equation}
\calP_{\mu\nu, \rho\sigma}=
\alpha\,I_{\mu\nu, \rho\sigma}+\beta\,\II_{\mu\nu, \rho\sigma}
+\gamma\,d_{\mu\nu}d_{\rho\sigma} + \delta\,(e_{\mu\nu}d_{\rho\sigma}
+d_{\mu\nu}e_{\rho\sigma}) +\varepsilon\,e_{\mu\nu}e_{\rho\sigma} ,
\end{equation}
is given by the inverse of the kinetic operator:
\begin{equation}
\calP_{\mu\nu, \alpha\beta}\,\calQ^{\alpha\beta}_{\ \ \ \rho\sigma}
=\frac12\left( \eta_{\mu\rho}\eta_{\nu\sigma}+ \eta_{\mu\sigma}\eta_{\nu\rho}\right)
= I_{\mu\nu, \rho\sigma}+\II_{\mu\nu, \rho\sigma}
+{1\over D-1}\,d_{\mu\nu}d_{\rho\sigma} + e_{\mu\nu}e_{\rho\sigma}.
\end{equation}
This condition requires
\begin{eqnarray}
&& \calA \alpha= 1, \quad
\calB \beta= 1, \quad
(D-1)\calC \gamma+ \calD \delta={1\over D-1},
\nn
&& (D-1)\calC \delta+ \calD \varepsilon=0,\quad
(D-1)\calD \gamma+ \calE \delta=0,\quad
(D-1)\calD \delta+ \calE \varepsilon=1,
\end{eqnarray}
so that we find
\begin{eqnarray}
&&\alpha={1\over\calA}= -{2\over p^2+m^2},\quad
\beta={1\over\calB}= -{2\alpha\over p^2+\alpha m^2},\quad \\
&&
\gamma=-{\calE\over(D-1)^2(\calD^2-\calE\calC)}
={2\over(D-1)(D-2)}\cdot{1\over p^2+2\beta wm^2}, \\
&&
\delta={\calD\over(D-1)(\calD^2-\calE\calC)}=
{4\beta x\over D-2}\cdot{1\over p^2+2\beta wm^2}, \\
&&
\varepsilon=-{\calC\over\calD^2-\calE\calC}
={8\beta^2x^2(D-1)\over D-2}\cdot{1\over p^2+2\beta wm^2}
-{4\alpha\beta^2\over p^2+\alpha\beta m^2}.
\end{eqnarray}
We see that most of the terms have gauge-dependent masses, which should cancel with
the Faddeev-Popov ghost.

\subsubsection{Faddeev-Popov ghost propagator}

The kinetic term of the Faddeev-Popov ghosts is:
\begin{equation}
i
\left(
\begin{array}{cc}
\bar c^\mu& \bar c
\end{array}
\right)
\left(
\begin{array}{cc}
-(p^2+\alpha m^2)d_{\mu\nu}-2y(p^2+\a\b m^2)e_{\mu\nu}
                         &  i m p_\mu(\a -2w)   \\
   0     & -p^2-  2\beta wm^2
\end{array}
\right)
\left(
\begin{array}{c}
c^\nu\\
c
\end{array}
\right) .
\end{equation}
We find the propagator from the inverse of this:
\begin{equation}
\bordermatrix{
 & c_\nu& c  \cr
\bar c_\mu
 & \displaystyle
  -{d_{\mu\nu}\over p^2+\alpha m^2}
  -\beta{e_{\mu\nu}\over p^2+\alpha\beta m^2}
 & \displaystyle
   {-ip_\mu\over p^2+\alpha m^2}(\alpha-2w){\b m\over p^2+2\beta wm^2} \cr
\bar c & 0 & \displaystyle
-{1\over p^2+2\beta wm^2} \cr} .
\end{equation}

\subsubsection{vector propagator}

The kinetic term of the vector field $A_\mu$ is given by
\begin{eqnarray}
&&-\frac12\left[ \left(p^2+\alpha m^2\right)\eta_{\mu\nu}-\left(1-\beta^{-1}\right)
p_\mu p_\nu\right] \nn
&&=
-\frac12\left[ \left(p^2+\alpha m^2\right)d_{\mu\nu}
+\beta^{-1}\left(p^2+\alpha\beta m^2\right)e_{\mu\nu}\right] ,
\end{eqnarray}
whose inverse gives the vector propagator:
\begin{eqnarray}
\langle A_\mu\,A_\nu\rangle&=&
-{d_{\mu\nu}\over p^2+\alpha m^2}-{\beta e_{\mu\nu}\over p^2+\alpha\beta m^2} \nn
&=&
{\eta_{\mu\nu}+{p_\mu p_\nu\over\alpha m^2}\over-p^2-\alpha m^2}
-\left({p_\mu p_\nu\over\alpha m^2}\right) {1\over-p^2-\alpha\beta m^2}.
\end{eqnarray}
Note that the massless singularities contained in
$d_{\mu\nu}$ and $e_{\mu\nu}$ have actually been cancelled in this
vector propagator. This should be so since
those singularities are of course not physical but an artifact of
our computational device using projection operators.
The same cancellations of massless singularities have
occurred also in the above tensor propagators, which the reader can confirm
using the above expressions for tensor propagator.

Summarizing, we have the following propagators after suitable normalization:

$h_{\mu\nu}$-sector:
\begin{equation}
\begin{array}{rll}
h_{\rm TT}: & \hbox{transverse-traceless $\frac{(D-2)(D+1)}{2}$-modes} &
\disp -{1\over p^2+m^2},  \\
h_{\rm LT}: & \hbox{longitudinal-transverse $(D-1)$-modes} &\disp -{1\over p^2+\alpha m^2},  \\
h_{\rm LL}+h: & \hbox{LL and trace $(1+1)$-modes} &\disp -{1\over p^2+\alpha\beta m^2},  \\
& &\disp -{1\over p^2+2\beta wm^2},
\end{array}
\end{equation}

$A_{\mu}$-$\pi$-sector:
\begin{equation}
\begin{array}{rll}
A_{\rm T}: & \hbox{massive vector $(D-1)$-modes} & \disp -{1\over p^2+\alpha m^2},  \\
A_{\rm L}: & \hbox{longitudinal 1-mode} &\disp -{1\over p^2+\alpha\beta m^2} , \\
 \pi: & \hbox{scalar 1-mode} &\disp -{1\over p^2+2\beta wm^2}.
\end{array}
\end{equation}

Faddeev-Popov ghost sector:
\begin{equation}
\begin{array}{rll}
\bar c_{\rm T},\ \ c_{\rm T}: & \hbox{massive $2(D-1)$-modes} &
\disp -{1\over p^2+\alpha m^2},  \\
\bar c_{\rm L},\ \ c_{\rm L}: & \hbox{longitudinal $(1+1)$-modes} &
\disp -{1\over p^2+\alpha\beta m^2} , \\
\bar c,\ \ c: & \hbox{scalar $(1+1)$-modes} & \disp -{1\over p^2+2\beta wm^2}.
\end{array}
\end{equation}

We see that almost all modes cancel out with the Faddeev-Popov ghosts,
and we are left with $\frac{(D-2)(D+1)}{2}$ (five for four dimensions) modes
of the symmetric transverse traceless tensor $h_{\mu\nu}$ with mass $m$.

\section{Absence of ghosts in the nonlinear massive gravity}
\label{full}

We now consider four-dimensional theory for the nonlinear massive gravity as formulated
by dRGT~\cite{DG,DGT1}.
For simplicity, here we discuss only four-dimensional theory, but the generalization to
arbitrary dimensions is straightforward.
The action is given by~\cite{DG,DGT1}
\bea
S = 
\int d^4 x \sqrt{- g} \Big[ R + m^2{\cal L}_{\rm mass} \Big],
\label{action1}
\ena
where ${\cal L}_{\rm mass}$ is given by
\bea
{\cal L}_{\rm mass} = \frac12 [(K_\mu^\mu)^2-K_\mu^\nu K_\nu^\mu]
+\frac{c_3}{3!} \e_{\mu\nu\rho\s} \e^{\a\b\c\s} K_\a^\mu K_\b^\nu K_\c^\rho
+\frac{c_4}{4!} \e_{\mu\nu\rho\s} \e^{\a\b\c\d} K_\a^\mu K_\b^\nu K_\c^\rho K_\d^\s.
\ena
Here $c_3, c_4$ are parameters and
\bea
{K^\mu}_\nu= \d^\mu{}_\nu-\c^\mu{}_\nu,~~~~
\c^\mu{}_\nu= \sqrt{g^{\mu\s} f_{\s\nu}},
\label{gamma}
\ena
where $f_{\mu\nu}$ is a fiducial metric which can be chosen to be flat metric $\eta_{\mu\nu}$.
Actually we would like to keep the general coordinate invariance by introducing St\"{u}ckelberg
field $Y^M$.
Following \cite{AGS}, we set the fiducial metric to
\bea
f_{\mu\nu} = \pa_\mu Y^M G_{MN} \pa_\nu Y^N.
\label{stuckel}
\ena
Here $Y^M$ is a coordinate in the ``target space'' and we can set it to
\bea
Y^M(x)=x^\mu\d_\mu^M +\phi^M(x),
\ena
obtaining
\bea
\pa_\mu Y^M=\d_\mu^M+\pa_\mu\phi^M.
\ena
where $\mu$ and $M$ represent the ``worldsheet'' and
``target space'' indices, respectively.
The original dRGT formulation corresponds to taking the target space metric
$G_{MN}$ flat Minkowski's $\eta_{MN}$ as we follow henceforth. We then have
\begin{equation}
f_{\mu\nu}= \eta_{\mu\nu} + (\partial_\mu\phi_\nu+\partial_\nu\phi_\mu)
 + \partial_\mu\phi^\rho\cdot \partial_\nu\phi_\rho,
\label{eq:fmunu}
\end{equation}
where we freely raise and lower the index $\mu$ of $\phi^\mu=\phi^M\delta^\mu_M$
by the Minkowski metric:
$\phi_\mu=\eta_{\mu\nu}\phi^\nu, \
\phi^\mu=\eta^{\mu\nu}\phi_\nu$.
We should note that though we use the formulation in which the general coordinate
invariance is recovered, our following discussions proceed with this choice of fiducial
metric; we restrict to the frame where the St\"{u}ckelberg fields $\phi_\mu$
have no vacuum expectation value.

\def\gbar{{\bar g}}
We are interested in the question whether there is higher time-derivative terms in
the St\"{u}ckelberg fields. To study this, we introduce the metric
fluctuation around general background $\gbar_{\mu\nu}$
\bea
g_{\mu\nu} = \gbar_{\mu\nu}+h_{\mu\nu}.
\ena
However, since we are interested only in the question whether there remains BD ghost
which exists in the St\"{u}ckelberg modes, we can
simply set the graviton fluctuation to zero
\begin{equation}
h_{\mu\nu}=0,
\end{equation}
and study the spectrum.
What we have to show now is that there is no higher derivative kinetic terms
for the St\"{u}ckelberg fields. If this is confirmed, the preceding discussion
shows that we have only five degrees of freedom and there is no BD ghost.

\subsection{Diagonalizing the background}
\label{dia}

\def\nmu{m}
\def\nnu{n}
\def\nrho{\ell}

The expansion of the square root $\sqrt{\gbar^{-1}f}$ around general background
$\gbar$ is very complicated in general if not impossible.
For example, one {\em cannot} simply make expansion like
\bea
\sqrt{A+B} \stackrel{?}{=} \sqrt{A} \Big(1+\frac12 A^{-1}B - \frac18 (A^{-1}B)^2 +\cdots\Big),
\ena
unless the matrices $A$ and $B$ commute with each other. We can make a general expansion
around a unit matrix as
\bea
\sqrt{A+B} = \sum_{n=0}^\infty{}_nC_{1/2} (A-1+B)^n,
\ena
with binomial coefficient $_nC_{1/2}$, but then the term $(A-1+B)^n$ is not so simple:
\bea
(A-1+B)^n = (A-1)^n + \sum_{k=1}^n (A-1)^{k-1}B(A-1)^{n-k} + O(B^2),
\ena
because $A$ and $B$ do not commute with each other in general. This expression is too
complicated to analyze. Our strategy is then to try to make the background diagonal, in which case
we can make more tractable expansion.

Consider the expression
\begin{equation}
\gbar^{\mu\rho}f_{\rho\nu}=
\gbar^{\mu\rho}\eta_{\rho\sigma}\Bigl(
\delta^\sigma{}_\nu+ \eta^{\sigma\tau}(\partial_\tau\phi_\nu+\partial_\nu\phi_\tau)
+ \eta^{\sigma\tau}\partial_\tau\phi^\alpha\cdot \partial_\nu\phi_\alpha
\Bigr),
\label{prod}
\end{equation}
or
\begin{equation}
\gbar^{-1}f =
(\gbar^{-1}\eta)
\Bigl(  1 + \eta^{-1}((\partial\phi)+ (\partial\phi)^T)
+ \eta^{-1}(\partial\phi)\eta^{-1}(\partial\phi)^T \Bigr),
\label{eq:gf}
\end{equation}
in matrix form.

The c-number part $\gbar^{-1}\eta$ can generally be made
diagonal by a matrix $V$.
This is true when all the eigenvectors of the $4\times4$ matrix $\gbar^{-1}\eta$
are independent and not degenerate. Degeneracy of the eigenvectors
may occur at measure-zero points in the functional space
of the background metric $\bar g_{\mu\nu}$. Moreover, as we shall see later
in an explicit example, we suspect that such a degeneracy occurs at
the metric $\gbar_{\mu\nu}$ which corresponds to rather singular and
unphysical background.
Therefore, we confine ourselves to the cases where the matrix $\gbar^{-1}\eta$
can be made diagonal.

Let $\alpha(\nnu)$ $(\nnu=1,\ 2,\ 3,\ 4)$ be the roots of the
characteristic equation $\det[x1- \gbar^{-1}\eta]=0$, and
$V_\nnu$ be eigenvectors of the matrix $\gbar^{-1}\eta$
belonging to the eigenvalue $\alpha(\nnu)$:
\begin{equation}
(\gbar^{-1}\eta)^\mu{}_\rho\ V^\rho{}_\nnu= \alpha(\nnu) V^\mu{}_\nnu
\quad \hbox{or}\quad (\gbar^{-1}\eta)\ V{}_\nnu= \alpha(\nnu) V{}_\nnu .
\label{eq:eigenEQ}
\end{equation}
Note that we use {\em roman letters} to denote the eigenvector labels in
distinction to the original vector indices denoted by Greek letters.
Since the matrix $\gbar^{-1}\eta$ satisfies
\begin{equation}
(\gbar^{-1}\eta)\, V = V \Bigl( \alpha(\nmu)\delta_{\nmu\nnu} \Bigr)
\qquad \hbox{for}\qquad V \equiv\Bigl(V_1,\ V_2,\ \cdots,\ V_4\Bigr),
\label{eq:V}
\end{equation}
it is made diagonal as
\begin{equation}
V^{-1}\,\gbar^{-1}\eta\, V = \Bigl( \alpha(\nmu)\delta_{\nmu\nnu} \Bigr) \equiv A^{(0)} .
\label{eq:317}
\end{equation}
Noting that $\gbar$ is real symmetric, we can show that the
matrix $V$ satisfies
\begin{equation}
V^{-1} = V^T \eta.
\label{eq:Vinv}
\end{equation}
Indeed, using Eq.~(\ref{eq:V}) and also its transpose, we can show
\begin{equation}
V^T (\eta\gbar^{-1}\eta)V
= V^T\eta V\, \Bigl( \alpha(\nmu)\delta_{\nmu\nnu} \Bigr)
= \Bigl( \alpha(\nmu)\delta_{\nmu\nnu} \Bigr)\,V^T\eta V\ .
\end{equation}
If all the eigenvalues are different one another, this implies
that $V^T\eta V$ is diagonal so that we can realize $V^T\eta V=1$ by
the normalization condition for the eigenvectors. Even if some eigenvalues
are degenerate, we can realize it as the ortho-normalization
condition in each common eigenvalue sector.

Performing the similarity transformation to (\ref{eq:gf}) by the matrix $V$,
and using the relation~\p{eq:Vinv}, we find
\begin{eqnarray}
V^{-1}(\gbar^{-1}f) V &=&
V^{-1}(\gbar^{-1}\eta) V \,V^{-1}\Bigl(  1 + \eta^{-1}((\partial\phi)+ (\partial\phi)^T)
+ \eta^{-1}(\partial\phi)\eta^{-1}(\partial\phi)^T \Bigr) V \nn
&=& A^{(0)}\,\Bigl(  1 + V^T\,((\partial\phi)+ (\partial\phi)^T)\,V
+ V^T\,(\partial\phi)\,V\,V^{T}\,(\partial\phi)^T\,V \Bigr) .
\end{eqnarray}
It is important to notice here that both the `vector' indices $\mu$ of $\partial_\mu$ and
of the St\"{u}ckelberg field $\phi_\mu$ are commonly transformed by the matrix $V$:
\begin{eqnarray}
&&\left[V^T\,(\partial\phi)\,V\right]_{\nmu\nnu}
=(V^T)_\nmu{}^\mu\,(\partial_\mu\phi_\nu)\,V^\nu{}_\nnu
= \bar\partial_\nmu\bar\phi_\nnu , \nn
&&\bar\partial_\nmu\equiv V^\mu{}_\nmu\,\partial_\mu, \quad
\bar\phi_\nmu\equiv\phi_\mu V^\mu{}_\nmu ,
\label{eq:bar}
\end{eqnarray}
so that
\begin{equation}
V^{-1}(\gbar^{-1}f) V
=A^{(0)}\,\Bigl(  1 + ((\bar\partial\bar\phi)+ (\bar\partial\bar\phi)^T)
+ (\bar\partial\bar\phi)(\bar\partial\bar\phi)^T \Bigr) .
\label{eq:simA}
\end{equation}
We should emphasize here that
the derivatives $\partial_\mu$ are only acting on the St\"{u}ckelberg
field and {\em never} differentiate the `rotation matrix' elements
$V^\mu{}_\nmu$ even if $V^\mu{}_\nmu$ are written after $\partial_\mu$.

Now the c-number part $A^{(0)}$ of this matrix is diagonal and its square root
is simply given by
\begin{equation}
\sqrt{A^{(0)}}\,{}_{\nmu\nnu}=
B^{(0)}{}_{\nmu\nnu}= \sqrt{\a(\nmu)} \,\delta_{\nmu\nnu}.
\label{eq:323}
\end{equation}
It is more convenient to make the matrix (\ref{eq:simA}) symmetric, so
we further perform the similarity transformation by $B^{(0)}$, and call the
resultant symmetric matrix $A$:
\begin{eqnarray}
A &\equiv& {B^{(0)}}^{-1}V^{-1} (\gbar^{-1}f) V\, B^{(0)} \nn
&=& B^{(0)}\,\Bigl(   1 + ((\bar\partial\bar\phi)+ (\bar\partial\bar\phi)^T)
+ (\bar\partial\bar\phi)(\bar\partial\bar\phi)^T \Bigr) B^{(0)} \nn
&\equiv& A^{(0)} + A^{(1)} + A^{(2)} .
\label{eq:A}
\end{eqnarray}
The matrices $A^{(1)}$ and $A^{(2)}$ are the linear and
quadratic terms, respectively, in the St\"{u}ckelberg field $\phi$ and
their matrix elements are given more explicitly by
\def\dbar#1{{\bar{\bar{#1}}}}
\begin{eqnarray}
&&A^{(1)}{}_{\nmu\nnu}= \sqrt{\alpha(\nmu)}(\bar\partial_\nmu\bar\phi_\nnu
+\bar\partial_\nnu\bar\phi_\nmu)\sqrt{\alpha(\nnu)}
= {\dbar\partial}_\nmu\dbar\phi_\nnu+\dbar\partial_\nnu\dbar\phi_\nmu, \nn
&& A^{(2)}{}_{\nmu\nnu}
= \sqrt{\alpha(\nmu)}\bar\partial_\nmu\bar\phi_\nrho\cdot\bar\partial_\nnu\bar\phi_\nrho\sqrt{\alpha(\nnu)}
= \dbar\partial_\nmu \bar\phi_\nrho\cdot\dbar\partial_\nnu\bar\phi_\nrho\ .
\label{eq:expnA}
\end{eqnarray}
Here the double barred quantities $\dbar\phi$ and $\dbar\partial$ are defined as
\begin{equation}
\dbar\phi_\nmu= \bar\phi_\nmu\sqrt{\alpha(\nmu)}
= \phi_\mu V^\mu{}_\nmu\sqrt{\alpha(\nmu)}= \phi_\mu(VB^{(0)})^\mu{}_\nmu ,
\end{equation}
and the same for $\dbar\partial_\nmu$ with understanding that the
derivative acts only on $\phi$ but neither on $\sqrt{\alpha(\nmu)}$
nor on
$V^\mu{}_\nmu$.
Remember that the `vector' index of the barred quantities $\bar\partial$ and $\bar\phi$
defined in (\ref{eq:bar}) now stands for the rotated one by $V$, and that of
double barred quantities $\dbar\partial$ and $\dbar\phi$ for the `rotated' one by
$V B^{(0)}$.

Before entering the detailed computation, let us look at the
`decoupling limit' at this stage.
Our inspection of the expressions (\ref{eq:expnA}) finds it natural to
define a decoupling limit by the following replacement similar to the decoupling limit
in the flat background case:
\begin{equation}
\bar\phi_\nmu\ \rightarrow\ \dbar\partial_\nmu\pi
\qquad \hbox{or, equivalently}\qquad
\phi_\mu\ \rightarrow\ (\partial_\nu\pi)(VB^{(0)}V^{-1})^\nu{}_\mu.
\label{eq:DecouplingLimit}
\end{equation}
It should be noted that the coefficients $(VB^{(0)}V^{-1})^\nu{}_\mu$ here
must be {\em real} in order for this replacement to make sense. This is because
$\phi_\mu$ and $\partial_\nu\pi$ are real fields. Fortunately, from (\ref{eq:317}) and
(\ref{eq:323}),
we have $VA^{(0)}V^{-1}=\gbar^{-1}\eta$ and hence
\begin{equation}
VB^{(0)}V^{-1}= V\sqrt{A^{(0)}}V^{-1}= \sqrt{VA^{(0)}V^{-1}}=\sqrt{\gbar^{-1}\eta}\ ,
\end{equation}
so that $VB^{(0)}V^{-1}$ is a real matrix
as long as $\sqrt{\gbar^{-1}\eta}$ is real. But the latter is the very
condition that the present dRGT theory has the hermitian mass term so
that it holds as long as the present theory makes sense.

We also note that this decoupling limit is quite nontrivial
because it mixes time and spatial derivatives by
the coefficients $(VB^{(0)}V^{-1})^\nu{}_\mu$ in general.
This happens when the background metric $\bg$ has
time-space component (shift).
We will see this in more detail in an explicit example later.
On the flat background $\gbar=\eta$, this of course reduces to the usual one
$\phi_\mu\rightarrow\partial_\mu\pi$ and do not mix them.

In this decoupling limit on general background, we have
\begin{eqnarray}
\dbar\partial_\nmu\bar\phi_\nnu&\rightarrow&
\dbar\partial_\nmu\dbar\partial_\nnu\pi
+ (\partial_\nu\pi)[\dbar\partial_\nmu(VB^{(0)}V^{-1})\cdot V]^\nu{}_\nnu, \nn
\dbar\partial_\nmu\dbar\phi_\nnu&\rightarrow&
(\dbar\partial_\nmu\dbar\partial_\nnu\pi)\sqrt{\alpha(\nnu)}
+ (\partial_\nu\pi)[\dbar\partial_\nmu(VB^{(0)}V^{-1})\cdot VB^{(0)}]^\nu{}_\nnu ,
\label{eq:Decoupling2}
\end{eqnarray}
with $\dbar\partial_\nmu\dbar\partial_\nnu\pi$ denoting
\begin{equation}
\dbar\partial_\nmu\dbar\partial_\nnu\pi
\equiv (\partial_\mu\partial_\nu\pi) (VB^{(0)})^\mu{}_\nmu
(VB^{(0)})^\nu{}_\nnu \,.
\label{eq:Decoupling3}
\end{equation}
That is, the derivative operator $\dbar\partial_\nmu$ here is understood to act
only on the field $\pi$ but not on the coefficients
$(VB^{(0)})^\nu{}_\nnu$, and
then $\dbar\partial_\nmu$ and $\dbar\partial_\nnu$ are commutative on $\pi$.
If we define a symmetric matrix $\Pi$ by
\begin{equation}
\Pi_{\nmu\nnu} =
\dbar\partial_\nmu\dbar\partial_\nnu\pi
\end{equation}
then, from
Eq.~(\ref{eq:expnA}),
we have in this limit
\begin{eqnarray}
&&A^{(1)}{}_{\nmu\nnu}\ \rightarrow\ (B^{(0)}\Pi+ \Pi B^{(0)})_{\nmu\nnu}
+ (\partial\pi\hbox{-term}), \nn
&& A^{(2)}{}_{\nmu\nnu}\ \rightarrow\ (\Pi^2)_{\nmu\nnu}
+ (\partial\pi\hbox{-term}),
\label{eq:expnA2}
\end{eqnarray}
where $(\partial\pi\hbox{-term})$ denotes the first order derivative terms of $
\pi$ field. Namely, if we keep only the second order derivative terms of
$\pi$ neglecting the first order derivative terms, then the matrix
$A$ takes very simple form:
\begin{eqnarray}
A
&=& A^{(0)} + A^{(1)} + A^{(2)} \nn
&=& (B^{(0)})^2 + (B^{(0)}\Pi+ \Pi B^{(0)})+ \Pi^2
= (B^{(0)}+\Pi)^2 .
\end{eqnarray}
That is, as far as the second order derivative terms $\partial\partial \pi$ are concerned,
\begin{equation}
B=\sqrt{A}= B^{(0)}+\Pi
\label{eq:declimit}
\end{equation}
in this decoupling limit and so there appear
no quadratic terms of the St\"{u}ckelberg field $\Pi$.
This is very similar situation to the flat background case, where actually
it gave dRGT the motivation for taking the square root form
for the mass term.
This form (\ref{eq:declimit}) of $\sqrt{A}$ guarantees that the
the dRGT mass terms generated by
$\det[1+\lambda\sqrt{A}]$ clearly have {\em total derivative} forms in the
decoupling limit as far as the higher
derivative terms $\partial\partial \pi$ are concerned.

Therefore, similarly to the flat case, we expect that
the original St\"{u}ckelberg `vector' field $\phi_\mu$ appears only in
the following `gauge invariant' tensor combination in the
quadratic terms in the mass term:
\begin{equation}
F_{\nmu\nnu} = \dbar\partial_\nmu\bar\phi_\nnu- \dbar\partial_\nnu\bar\phi_\nmu.
\end{equation}
This combination of $\dbar\partial$ and $\bar\phi$ is suitable because of the form
(\ref{eq:DecouplingLimit})
of the decoupling limit $\bar\phi_\mu\ \rightarrow\ \dbar\partial_\mu\pi$.
We shall now show that this is indeed the case
if we neglect some lower order
derivative terms.

\subsection{Computing the general mass terms}
\label{mass}

Let us compute the generating function of the general
mass terms:
\begin{equation}
\det[ 1+ \lambda\sqrt{\gbar^{-1}f} ] .
\end{equation}
Since this is invariant under the similarity transformation, we can use
the expression $A$ in \p{eq:A} for the matrix $\sqrt{\bg^{-1}f}$:
\begin{equation}
\det[ 1+ \lambda\sqrt{\gbar^{-1}f} ]
=\det[ {B^{(0)}}^{-1}V^{-1}(1+ \lambda\sqrt{\gbar^{-1}f})V B^{(0)} ]
=\det[ 1+ \lambda\sqrt{A\,} \,]
\end{equation}
The square root of the matrix $A$ can be calculated order by order in
the St\"{u}ckelberg field $\phi$
thanks to the fact that the matrix $B^{(0)}$ is diagonal.
The matrix equation
\begin{equation}
B^{(0)}* X \equiv B^{(0)}X + XB^{(0)} = C,
\end{equation}
for $X$ can be solved explicitly~\cite{GLM}. The solution $X$ to this equation,
denoted formally as
$(B^{(0)}*)^{-1} C$,
is given explicitly by
\begin{equation}
X_{\nmu\nnu}= \left((B^{(0)}*)^{-1} C\right)_{\nmu\nnu}
=\frac1{\sqrt{\a(\nmu)}+\sqrt{\a(\nnu)}} C_{\nmu\nnu}.
\end{equation}
This formula enables us to find the square root of $A$:
\bea
\sqrt{A^{(0)} + A^{(1)} + A^{(2)}}\,{}_{\nmu\nnu}= B^{(0)}{}_{\nmu\nnu}
+ B^{(1)}{}_{\nmu\nnu}+B^{(2)}{}_{\nmu\nnu} + \cdots,
\ena
with
\bea
&& B^{(1)}{}_{\nmu\nnu}= \frac1{\sqrt{\a(\nmu)}+\sqrt{\a(\nnu)}} A^{(1)}{}_{\nmu\nnu}, \nn
&& B^{(2)}{}_{\nmu\nnu}= \frac1{\sqrt{\a(\nmu)}+\sqrt{\a(\nnu)}} (A^{(2)}{}_{\nmu\nnu}
-(B^{(1)}B^{(1)}){}_{\nmu\nnu}).
\ena
Substituting the expression (\ref{eq:expnA}), we find
\begin{eqnarray}
&& B^{(1)}{}_{\nmu\nnu}
=\frac1{\sqrt{\alpha(\nmu)}+\sqrt{\alpha(\nnu)}} \left({\dbar\partial}
\dbar\phi\right)_{(\nmu\nnu)}, \nn
&& B^{(2)}{}_{\nmu\nnu}= \frac1{\sqrt{\alpha(\nmu)}+\sqrt{\alpha(\nnu)}} \sum_\nrho
\biggl\{\dbar\partial_\nmu\bar\phi_\nrho\cdot\dbar\partial_\nnu\bar\phi_\nrho\nn
&&\hspace{5eM}-\frac1{(\sqrt{\alpha(\nmu)}+\sqrt{\alpha(\nrho)})(\sqrt{\alpha(\nnu)}
+\sqrt{\alpha(\nrho)})}
\Bigl({\dbar\partial}\dbar\phi\Bigr)_{(\nmu\nrho)}
\Bigl({\dbar\partial}\dbar\phi\Bigr)_{(\nnu\nrho)} \biggr\},
\end{eqnarray}
with notation
$\bigl({\dbar\partial}\dbar\phi\bigr)_{(\nmu\nnu)}
\equiv\dbar\partial_\nmu\dbar\phi_\nnu+\dbar\partial_\nnu\dbar\phi_\nmu$.

Now we expand the determinant $\det[ 1+ \lambda\sqrt{A\,} \,]=\det[ 1+ \lambda B\,]$
in powers of the St\"{u}ckelberg field $\phi$:
\begin{eqnarray}
\det[ 1+ \lambda B\,] &=&
\det[ 1+ \lambda(B^{(0)} + B^{(1)} + B^{(2)}) \,] \nn
&=&\det[ 1+ \lambda B^{(0)}\,] \cdot
\det\left[ 1+ \beta^{(1)} + \beta^{(2)}\right], \nn
&& \beta^{(n)} \equiv\frac{\lambda}{1+\lambda B^{(0)}}B^{(n)}, \quad (n=1,2).
\end{eqnarray}
The quadratic terms in $\phi$ is thus given by
\begin{eqnarray}
\det[ 1+ \lambda B\,]\Bigr|_{\rm quad}=\det[ 1+ \lambda B^{(0)}\,] \cdot
\left\{
\tr \bigl[ \beta^{(2)}\bigr]
+ {1\over2}\left(
\bigl( \tr\bigl[\beta^{(1)} \bigr] \bigr)^2
- \tr\bigl[ (\beta^{(1)})^2 \bigr]
\right)
\right\}.
\label{eq:qua}
\end{eqnarray}
We now simplify each term. First consider
\begin{eqnarray}
\tr \bigl[ \beta^{(2)}\bigr]
\!\!&=&\!\!
\sum_\nmu\frac{\lambda}{1+\lambda\sqrt{\alpha(\nmu)}}
\frac1{2\sqrt{\alpha(\nmu)}}\sum_{\nnu}
\left\{
\bigl({\dbar\partial}_\nmu\bar\phi_\nnu\bigr)^2 -
\frac1{(\sqrt{\alpha(\nmu)}+\sqrt{\alpha(\nnu)})^2}
\left({\dbar\partial}\dbar\phi\right)_{(\nmu\nnu)}^2
\right\} \nn
\!\!&=&\!\!
\frac12 \sum_{\nmu,\nnu}
\frac{\lambda}{1+\lambda\sqrt{\alpha(\nmu)}}
\frac{\bigl({\dbar\partial}_\nmu\bar\phi_\nnu-{\dbar\partial}_\nnu\bar\phi_\nmu\bigr)}
{\sqrt{\alpha(\nmu)}+\sqrt{\alpha(\nnu)}}
\left(
2{\dbar\partial}_\nmu\bar\phi_\nnu
-\frac{\sqrt{\alpha(\nmu)}\bigl({\dbar\partial}_\nmu\bar\phi_\nnu
-{\dbar\partial}_\nnu\bar\phi_\nmu\bigr)}
{\sqrt{\alpha(\nmu)}+\sqrt{\alpha(\nnu)}}
\right). \nn
\end{eqnarray}
Averaging with the term obtained by exchanging the dummy indices
$\nmu\leftrightarrow \nnu$, we get
\begin{eqnarray}
\tr \bigl[ \beta^{(2)}\bigr]
\!\!&=&\!\!
\frac14 \sum_{\nmu,\nnu}
\frac{\lambda}{(1+\lambda\sqrt{\alpha(\nmu)})(1+\lambda\sqrt{\alpha(\nnu)})}
\frac{\bigl({\dbar\partial}_\nmu\bar\phi_\nnu-{\dbar\partial}_\nnu\bar\phi_\nmu\bigr)}
{\sqrt{\alpha(\nmu)}+\sqrt{\alpha(\nnu)}} \nn
&& \hs{-5}\times\left\{
\left(
1-\frac{2\lambda\sqrt{\alpha(\nmu)\alpha(\nnu)}}
{\sqrt{\alpha(\nmu)}+\sqrt{\alpha(\nnu)}}\right)
\bigl({\dbar\partial}_\nmu\bar\phi_\nnu-{\dbar\partial}_\nnu\bar\phi_\nmu\bigr)
+2\lambda
\bigl({\dbar\partial}_\nmu\dbar\phi_\nnu-{\dbar\partial}_\nnu\dbar\phi_\nmu\bigr)
\right\}.~~~~
\label{eq:B2}
\end{eqnarray}
The contribution of the second term in the bracket here is combined with
the $\tr[\beta^{(1)}\beta^{(1)}]$ term to yield
\begin{eqnarray}
&&\hspace{-3em}-\frac12 \tr\bigl[ (\beta^{(1)})^2 \bigr]
+\hbox{(second term of Eq.~(\ref{eq:B2}))} \nn
&&=
-\frac12\sum_{\nmu,\nnu}
\frac{\lambda^2}{(1+\lambda\sqrt{\alpha(\nmu)})(1+\lambda\sqrt{\alpha(\nnu)})}
\frac1{(\sqrt{\alpha(\nmu)}+\sqrt{\alpha(\nnu)})^2} \times\nn
&&\qquad \times\left\{
\bigl({\dbar\partial}_\nmu\dbar\phi_\nnu+{\dbar\partial}_\nnu\dbar\phi_\nmu\bigr)^2
-(\sqrt{\alpha(\nmu)}+\sqrt{\alpha(\nnu)})
\bigl({\dbar\partial}_\nmu\bar\phi_\nnu-{\dbar\partial}_\nnu\bar\phi_\nmu\bigr)
\bigl({\dbar\partial}_\nmu\dbar\phi_\nnu-{\dbar\partial}_\nnu\dbar\phi_\nmu\bigr)
\right\} \nn
&&=
\frac12\sum_{\nmu,\nnu}
\frac{\lambda^2}{(1+\lambda\sqrt{\alpha(\nmu)})(1+\lambda\sqrt{\alpha(\nnu)})}
\frac1{(\sqrt{\alpha(\nmu)}+\sqrt{\alpha(\nnu)})^2} \times\nn
&&\qquad \times\left\{
\sqrt{\alpha(\nmu)\alpha(\nnu)}
\bigl({\dbar\partial}_\nmu\bar\phi_\nnu-{\dbar\partial}_\nnu\bar\phi_\nmu\bigr)^2
-(\sqrt{\alpha(\nmu)}+\sqrt{\alpha(\nnu)})^2
{\dbar\partial}_\nmu\bar\phi_\nnu\cdot{\dbar\partial}_\nnu\bar\phi_\nmu
\right\},
\end{eqnarray}
which cancels partially the first term in \p{eq:B2}. We are thus left with
\begin{eqnarray}
&&\hspace{-2em}\tr \bigl[ \beta^{(2)}\bigr]
-\frac12 \tr\bigl[ (\beta^{(1)})^2 \bigr]
=
\sum_{\nmu,\nnu}
\frac1{(1+\lambda\sqrt{\alpha(\nmu)})(1+\lambda\sqrt{\alpha(\nnu)})} \nn
&&\qquad \times
\left\{
\frac{\lambda}4\frac1{\sqrt{\alpha(\nmu)}+\sqrt{\alpha(\nnu)}}
\bigl({\dbar\partial}_\nmu\bar\phi_\nnu-{\dbar\partial}_\nnu\bar\phi_\nmu\bigr)^2
-\frac{\lambda^2}2
{\dbar\partial}_\nmu\bar\phi_\nnu\cdot{\dbar\partial}_\nnu\bar\phi_\nmu
\right\}.
\label{eq:334}
\end{eqnarray}
The first term takes a ``gauge-invariant" form while the second term is not.
The latter term is however almost ``cancelled" by the remaining term in \p{eq:qua}:
\begin{eqnarray}
+ {1\over2}
\bigl( \tr\bigl[\beta^{(1)} \bigr] \bigr)^2
= \frac{\lambda^2}2
\sum_{\nmu,\nnu}
\frac1{(1+\lambda\sqrt{\alpha(\nmu)})(1+\lambda\sqrt{\alpha(\nnu)})}
\left(
{\dbar\partial}_\nmu\bar\phi_\nmu\cdot{\dbar\partial}_\nnu\bar\phi_\nnu
\right).
\end{eqnarray}
If we could do partial integration with respect to the differential operators
${\dbar\partial}_\nmu$ and ${\dbar\partial}_\nnu$ here, this term would really cancel the second
term in (\ref{eq:334}). But, there are various $x$-dependent factors
$\sqrt{\alpha(\nmu)}$'s and $V^\mu{}_\nmu$
in front of the differential operators, the cancellation
is not complete, and the terms with lower derivative terms of the form
$\phi\partial\phi$ or $\phi\phi $ remain.

The final quadratic terms are thus given by
\begin{eqnarray}
&&\hspace{-2em}
\det[ 1+ \lambda B\,]\Bigr|_{\rm quad}
=
\prod_\ell (1 + \lambda\sqrt{\alpha(\ell)}) \cdot
\sum_{\nmu,\nnu}
\frac1{(1+\lambda\sqrt{\alpha(\nmu)})(1+\lambda\sqrt{\alpha(\nnu)})}
\nn
&&\qquad \times
\left\{
\frac\lambda{\sqrt{\alpha(\nmu)}+\sqrt{\alpha(\nnu)}}\frac14
\bigl({\dbar\partial}_\nmu\bar\phi_\nnu-{\dbar\partial}_\nnu\bar\phi_\nmu\bigr)^2
+\frac{\lambda^2}2
\left(
{\dbar\partial}_\nmu\bar\phi_\nmu\cdot{\dbar\partial}_\nnu\bar\phi_\nnu
-{\dbar\partial}_\nmu\bar\phi_\nnu\cdot{\dbar\partial}_\nnu\bar\phi_\nmu
\right)
\right\}.\nn
\label{eq:344}
\end{eqnarray}

\subsection{Gauge invariance and the no-ghost theorem}
\label{hidden}

As anticipated from the consideration of the decoupling limit,
the resultant generic mass term is almost
``gauge invariant" under
\begin{equation}
\delta\bar\phi_\nmu= \dbar\partial_\nmu\Lambda, \quad \hbox{or, more precisely,}\quad
\delta\phi_\mu= (\partial_\nu\Lambda)(VB^{(0)}V^{-1})^\nu{}_\mu.
\label{eq:351}
\end{equation}
Actually it is not exactly invariant since
the coefficients $V^\mu{}_\nmu$ and $\sqrt{\alpha(\nmu)}$ are $x$-dependent and
the derivatives do not commute with them.
So we find that it is convenient to introduce the St\"{u}ckelberg `scalar' field $\pi$ by
\begin{equation}
\phi_\mu= A_\mu+ (\partial_\nu\pi)(VB^{(0)}V^{-1})^\nu{}_\mu.
\label{eq:Stuckelberg2}
\end{equation}
Then the $U(1)$ gauge invariance under
\begin{equation}
\delta A_\mu= (\partial_\nu\Lambda)(VB^{(0)}V^{-1})^\nu{}_\mu
\quad \hbox{and}\quad
\delta\pi= -\Lambda,
\end{equation}
becomes exact since the change cancels between $A_\mu$ and $\partial\pi$ terms
leaving $\phi_\mu$ intact. It is important to make this U(1) gauge invariance
exact; this is because it is lifted to the BRST invariance to define
the physical subspace in covariant gauges so that it must be an {\em exact
gauge symmetry} of the total action.

The above mentioned {\em approximate `gauge invariance'} under (\ref{eq:351}),
on
the other hand, guarantees that the
higher derivative terms in the kinetic term of the $\pi$-field cancel.
This is essentially due to the fact that the St\"{u}ckelberg field expression
(\ref{eq:Stuckelberg2}) for $\phi_\mu$ is defined in accordance with
the decoupling limit (\ref{eq:DecouplingLimit}).

Let us now explicitly show that the higher derivative terms of the $\pi$-field
indeed cancel in the kinetic term (\ref{eq:344}).

First, consider the first term in (\ref{eq:344}) written in terms of
$
F_{\nmu\nnu} = \dbar\partial_\nmu\bar\phi_\nnu- \dbar\partial_\nnu\bar\phi_\nmu.
$
Note that the St\"{u}ckelberg expression (\ref{eq:Stuckelberg2})
for $\phi$ gives
\begin{equation}
\dbar\partial_\nmu\bar\phi_\nnu=
\dbar\partial_\nmu\bar A_\nnu+
\dbar\partial_\nmu\dbar\partial_\nnu\pi+ C^\rho_{\nmu\nnu}\partial_\rho\pi,
\end{equation}
where $\dbar\partial_\nmu\dbar\partial_\nnu\pi$ is defined in (\ref{eq:Decoupling3}) and
the coefficient $C^\rho_{\nmu\nnu}$ of $\partial\pi$ term is given by
\begin{equation}
C^\rho_{\nmu\nnu} = [\dbar\partial_\nmu(VB^{(0)}V^{-1})\cdot V]^\rho{}_\nnu.
\end{equation}
Recalling that $\dbar\partial_\nmu\dbar\partial_\nnu\pi$
defined in (\ref{eq:Decoupling3}) is symmetric under
$\nmu \leftrightarrow \nnu$,
we see that the second order derivative terms $\dbar\partial\dbar\partial\pi$ cancel in
\begin{equation}
F_{\nmu\nnu} \equiv
{\dbar\partial}_\nmu\bar\phi_\nnu-{\dbar\partial}_\nnu\bar\phi_\nmu
=({\dbar\partial}_\nmu\bar A_\nnu-{\dbar\partial}_\nnu\bar A_\nmu)
+ (C^\rho_{\nmu\nnu}-C^\rho_{\nnu\nmu})\partial_\rho\pi,
\end{equation}
so that the first term in (\ref{eq:344}) contains only the
first order derivative $\partial\pi$ of the $\pi$ field.

Next, consider the second term in (\ref{eq:344}).
In order to do the partial integration carefully, let us make
explicit the factors contained in the
definitions of barred quantities:
\begin{equation}
\bar X_\nmu= X_\mu V^\mu{}_\nmu,\qquad
\dbar X_\nmu= X_\mu(VB^{(0)})^\mu{}_\nmu.
\end{equation}
We define the coefficient $C^{\mu\nu}$ which will
frequently appear below:
\begin{equation}
C^{\mu\nu}\equiv(VB^{(0)}V^T)^{\mu\nu}=C^{\nu\mu}.
\end{equation}
Noting $V^{-1}=V^T\eta$, we can rewrite the St\"{u}ckelberg field expression
(\ref{eq:Stuckelberg2}) in the form
\begin{equation}
\phi_\mu=A_\mu+(\partial_\nu\pi)(VB^{(0)}V^T)^{\nu\rho}\eta_{\rho\mu}
= A_\mu+\eta_{\mu\rho}C^{\rho\nu}\partial_\nu\pi.
\end{equation}
We find
\begin{equation}
{\dbar\partial}_\nmu\bar\phi_\nmu= \partial_\rho\phi_\sigma(VB^{(0)})^\rho{}_\nmu V^\sigma{}_\nmu
= \partial_\rho\phi_\sigma(VB^{(0)}V^T)^{\rho\sigma}
= C^{\rho\sigma}\partial_\rho\phi_\sigma,
\end{equation}
and, similarly,
\begin{eqnarray}
{\dbar\partial}_\nmu\bar\phi_\nmu\cdot{\dbar\partial}_\nnu\bar\phi_\nnu
-{\dbar\partial}_\nmu\bar\phi_\nnu\cdot{\dbar\partial}_\nnu\bar\phi_\nmu
=C^{\mu\alpha}C^{\nu\beta}
\bigl(\partial_\mu\phi_\alpha\cdot\partial_\nu\phi_\beta
-\partial_\mu\phi_\beta\cdot\partial_\nu\phi_\alpha\bigr).
\end{eqnarray}
Consequently the second term in (\ref{eq:334}) can be put, after performing
partial integrations twice, into the form
\begin{equation}
c \bigl({\dbar\partial}_\nmu\bar\phi_\nmu\cdot{\dbar\partial}_\nnu\bar\phi_\nnu
-{\dbar\partial}_\nmu\bar\phi_\nnu\cdot{\dbar\partial}_\nnu\bar\phi_\nmu\bigr)
=\phi_\alpha\,\partial_\mu\partial_\nu(cC^{\mu\alpha}C^{\nu\beta})\cdot\phi_\beta
+2\phi_\alpha\,\partial_\nu(cC^{\mu\alpha}C^{\nu\beta})\cdot\partial_\mu\phi_\beta,
\label{eq:355}
\end{equation}
where $c$ stands for all the prefactors
in front of this term in the action (including $\det\sqrt{\gbar}$).
Now the first term on the right hand side of (\ref{eq:355}) contains
only $\phi$'s with no derivatives so that it contains at most first order
derivatives of $\pi$-fields. The second term looks containing $\partial\phi$ which
gives second order derivative of $\pi$ since
\begin{equation}
\partial_\mu\phi_\beta
= \partial_\mu A_\beta+ \partial_\mu(\eta_{\beta\rho}C^{\rho\nu}\partial_\nu\pi).
\label{eq:356}
\end{equation}
Nevertheless we now show that those second order derivative terms of $\pi$ vanish.
Since the first order derivative of the `vector'
field $A_\mu$ is in any case contained in the action, we can forget about it here.
Keeping only the $\pi$ field in $\phi$, we find that
the second term of (\ref{eq:355}) becomes
\begin{eqnarray}
&&\hspace{-2em}2\phi_\alpha\,\partial_\nu(cC^{\mu\alpha}C^{\nu\beta})\cdot\partial_\mu
\phi_\beta\Big|_{\pi^2\ {\rm terms}}\nn
&=&
2\eta_{\alpha\delta}C^{\delta\rho}\partial_\rho\pi\cdot \partial_\nu(cC^{\mu\alpha}C^{\nu\beta})\cdot
\Bigl(
\eta_{\beta\gamma}\partial_\mu C^{\gamma\tau}\cdot\partial_\tau\pi+
\eta_{\beta\gamma}C^{\gamma\tau}\cdot\partial_\mu\partial_\tau\pi\Bigr).
\label{eq:357}
\end{eqnarray}
The first term is harmless with only the first derivatives on $\pi$'s,
but the last term is the dangerous one containing the second derivative $\partial\partial \pi$
which we write in the form
\begin{eqnarray}
\hbox{The last term of (\ref{eq:357})}
=
2d^{\mu\nu\rho}\partial_\nu\pi\cdot \partial_\mu\partial_\rho\pi\equiv L,
\label{eq:359}
\end{eqnarray}
by introducing a coefficient
\begin{equation}
d^{\mu\nu\rho}\equiv
(\eta C)_\alpha{}^\nu
(\eta C)_\beta{}^\rho\partial_\gamma(cC^{\mu\alpha}C^{\gamma\beta}).
\end{equation}
By performing partial integration for $\partial_\mu$,
we can rewrite (\ref{eq:359}) as
\begin{eqnarray}
L &=&
-2d^{\mu\nu\rho}\partial_\mu\partial_\nu\pi\cdot \partial_\rho\pi
-2(\partial_\mu d^{\mu\nu\rho})\partial_\nu\pi\cdot \partial_\rho\pi.
\label{eq:360}
\end{eqnarray}
Averaging these two expressions (\ref{eq:359}) and (\ref{eq:360}), we have
\begin{equation}
L =
(d^{\mu\rho\nu}-d^{\mu\nu\rho})\partial_\mu\partial_\nu\pi\cdot \partial_\rho\pi
-(\partial_\mu d^{\mu\nu\rho})\partial_\nu\pi\cdot \partial_\rho\pi.
\end{equation}
Noticing that the coefficient of the first term
$d^{\mu\rho\nu}-d^{\mu\nu\rho}\equiv2d^{\mu[\rho\nu]}$ is antisymmetric
under $\rho\leftrightarrow \nu$, we can make a partial integration to put it into the
first order derivative terms:
\begin{eqnarray}
L &=&
-2(\partial_\nu d^{\mu[\rho\nu]})\partial_\mu\pi\cdot \partial_\rho\pi
-(\partial_\mu d^{\mu\nu\rho})\partial_\nu\pi\cdot \partial_\rho\pi\nn
&=&
\Bigl(\partial_\nu\left(d^{\mu\nu\rho}-d^{\mu\rho\nu}-d^{\nu\mu\rho}\right)\Bigr)
\partial_\mu\pi\cdot \partial_\rho\pi.
\end{eqnarray}
We have thus shown that all the $\pi$ field terms can be put solely into
first order derivative terms. So the quadratic part in fields of the mass
term takes the usual form $L(\varphi, \partial\varphi )$ containing only up to
first order derivatives for all the fields $\varphi = \{h_{\mu\nu}, A_\mu, \pi\}$.
\footnote{Although we have set $h_{\mu\nu}=0$ in this calculation, it is clear
that $h_{\mu\nu}$ appears only without derivatives in the mass term, so that
it can appear in the quadratic term in the form $h \partial\phi$ at the highest
derivative order. $h \partial\phi$ contains the second order derivative of $\pi$,
$h\partial\partial \pi$, but it can be rewritten into the first order derivative term
$\partial h\cdot \partial\pi$.}

The $U(1)$ gauge invariance is exact and all the fields appear only up to the
first order derivative in the quadratic kinetic term. On any background metric,
the particle modes are determined by the quadratic terms.
Combined with our previous counting of physical degrees of freedom,
this implies that there appears no BD ghost mode in this theory on the general
background metric, and completes our proof of no-ghost theorem.

\section{Discussions}

It is instructive to see the general result in the previous section
explicitly for a concrete nontrivial background example. Let us consider
the following background metric $\gbar_{\mu\nu}$ discussed by dRGT~\cite{DGT2}:
\begin{equation}
ds^2 = \gbar_{\mu\nu}dx^\mu dx^\nu=-dt^2 + \delta_{ij}(dx^i + 2l^idt)(dx^j + 2l^jdt).
\end{equation}
This is the metric with the lapse $N=1$ and the shift vector
$N^i=2l^i$. Since the space metric $\gamma_{ij}$ is taken to be
$\delta_{ij}$, we can freely
rotate the spatial axis such that the shift vector points the $x^1$ direction:
\begin{equation}
\delta_{ij}l^idx^j =  l dx^1\ .
\end{equation}
For this background metric $\gbar_{\mu\nu}$, we have
\begin{equation}
(\gbar^{-1}\eta)^\mu{}_\nu=
\begin{pmatrix}
1 & 2l && \\
-2l & 1-4l^2 && \\
&& 1 & \\
&&&1 \\
\end{pmatrix} ,
\label{eq:gbareta}
\end{equation}
where the blank entry is all zero.
The characteristic equation for the first nontrivial $2\times2$ matrix
in the $(x^0,\,x^1)$ subspace is
\begin{equation}
x^2-2(1-2l^2)x+1=0.
\label{eq:character}
\end{equation}
The metric is flat for $l=0$. For the reason to be clear shortly, we consider
only the case $|l|<1$. The eigenvalues are then complex:
\begin{equation}
\begin{cases}
\alpha(0) = \alpha\\ \alpha(1)= \alpha^*
\end{cases}
\qquad \hbox{with}\qquad
\alpha= (\sqrt{1-l^2} + il)^2.
\end{equation}
The eigenvectors for these two eigenvalues
in the $(x^0,x^1)$ subspace are conveniently chosen as
\begin{equation}
V_{(2)}= (V_1,\ V_2) = {1\over N}
\begin{pmatrix}
-ia^* & ia \\
a & a^* \\
\end{pmatrix}
\qquad \hbox{with}\qquad
\begin{array}{l}
a\equiv\sqrt{\sqrt{1-l^2} + il} =\sqrt[4]{\alpha} \\
N\equiv\sqrt2 \sqrt[4]{1-l^2} \\
\end{array}.
\end{equation}
Note that $\alpha$ and $a$ are unimodular: $\alpha\alpha ^*=1 = aa^*$, and satisfy
\begin{equation}
\sqrt{\alpha}+\sqrt{\alpha^*}=2\sqrt{1-l^2},\quad  i(\sqrt{\alpha}-\sqrt{\alpha^*})=-2l,\quad
\alpha+\alpha^* = 2(1-2l^2).
\label{eq:407}
\end{equation}
The other two eigenvalues and eigenvectors in the $(x^2, x^3)$ directions are trivial.
Hence the matrix $V$ which diagonalizes the matrix $\gbar^{-1}\eta$
in (\ref{eq:gbareta}) and the diagonalized matrix are given by
\begin{equation}
V=
\begin{pmatrix}
V_{(2)}& \\
& 1_2 \\
\end{pmatrix}
\quad \rightarrow\quad V^{-1} \gbar^{-1}\eta V =
\begin{pmatrix}
\alpha&&& \\
& \alpha^*&& \\
&&1&\\
&&&1\\
\end{pmatrix}.
\end{equation}
Note that this matrix $V$ is properly normalized so as to satisfy
Eq.~(\ref{eq:Vinv}):
\begin{equation}
V^{-1}=V^T\eta.
\end{equation}

Now the barred derivatives $\bar\partial_m= V^\mu{}_m\partial_\mu$
defined in Eq.~(\ref{eq:bar}) are explicitly read as
\begin{eqnarray}
\bar\partial_0 &=& \frac1N\left( -ia^*\partial_0+a\partial_1\right), \qquad \dbar\partial_0
= \sqrt{\alpha}\bar\partial_0, \nn
\bar\partial_1 &=& \frac1N\left( ia\partial_0+a^*\partial_1\right)=\bar\partial_0^*,
\qquad \dbar\partial_1= \sqrt{\alpha^*}\bar\partial_1=\dbar\partial_0^*,
\label{eq:409}
\end{eqnarray}
and, $\bar\partial_2=\partial_2,\ \bar\partial_3=\partial_3$, of course.
The barred fields $\bar\phi_m=V^\mu{}_m\phi_\mu$ are similar:
\begin{eqnarray}
\bar\phi_0 &=& \frac1N\left( -ia^*\phi_0+a\phi_1\right), \qquad
\dbar\phi_0= \sqrt{\alpha}\bar\phi_0, \nn
\bar\phi_1 &=& \frac1N\left( ia\phi_0+a^*\phi_1\right)=\bar\phi_0^*,
\qquad \dbar\phi_1= \sqrt{\alpha^*}\bar\phi_1=\dbar\phi_0^*.
\label{eq:410}
\end{eqnarray}

Our result for the general mass term $\det[1+\lambda B]|_{\rm quad}$
was given in Eq.~(\ref{eq:344}). If we keep only the nontrivial terms
$\bar\partial_m\bar\phi_n$ with
$(m,n)=(1,0)$ and $(0,1)$, it gives
\begin{eqnarray}
\det[1+\lambda B]|_{\rm quad}
&=&
(1 + \lambda)^2
\left\{
\frac\lambda{\sqrt{\alpha}+\sqrt{\alpha^*}}\frac12
\bigl(\sqrt{\alpha}{\bar\partial}_0\bar\phi_1-\sqrt{\alpha^*}{\bar\partial}_1\bar\phi_0\bigr)^2
\right.
\nn
&&{}\hspace{5em}
+{\lambda^2}\alpha\alpha ^*
\left(
{\bar\partial}_0\bar\phi_0\cdot{\bar\partial}_1\bar\phi_1
-{\bar\partial}_0\bar\phi_1\cdot{\bar\partial}_1\bar\phi_0
\right)
\biggr\}.
\end{eqnarray}
Substituting Eqs.~(\ref{eq:409}) and (\ref{eq:410}), and using
Eq.~(\ref{eq:407}), we find that this reduces to
\begin{eqnarray}
(1 + \lambda)^2
\left\{
\frac\lambda{4(1-l^2)^{3/2}}
\bigl(\dot\phi_1-l\dot\phi_0 + (2l^2-1)\phi'_0-l\phi'_1\bigr)^2
+{\lambda^2}
\left( \dot\phi_1\phi'_0-\dot\phi_0\phi'_1
\right)
\right\}.
\label{eq:413}
\end{eqnarray}
where $\dot\phi\equiv\partial_0\phi,\ \phi'\equiv\partial_1\phi$.
Note that the second term has lost the $x^\mu$-dependent coefficients and
the overall factor $\sqrt{-\gbar}=1$ in front is also $x^\mu$-independent. So
the second term can be partial-integrated away.
Note also that the first term contains the square of $\dot\phi_0$, which would
yield the square of the second order time derivative $\ddot\pi$ if we had
introduced the St\"{u}ckelberg scalar field $\pi$ in the same manner
as the flat background case:
\begin{equation}
\phi_\mu\rightarrow\partial_\mu\pi.
\label{eq:flats}
\end{equation}
As was argued in Ref.~\cite{DGT2}, this term is actually harmless because $\dot\phi_0$ comes into
the action only with the particular combination
($\dot\phi_1-l\dot\phi_0$) with $\dot\phi_1$
and does not give rise
to another degree of freedom than $\phi_1$. In our discussions, we can see the absence of
ghost in a better way.
It is important to remember that the proper way of introducing $\pi$ in the general
background is not \p{eq:flats} but
\begin{equation}
\phi_\mu\rightarrow\partial_\nu\pi(VB^{(0)}V^{-1})^\nu{}_\mu,
\end{equation}
as given in Eq.~(\ref{eq:Stuckelberg2}). The coefficient $(VB^{(0)}V^{-1})^\nu{}_\mu$ reads
\begin{equation}
VB^{(0)}V^{-1}= \sqrt{\gbar^{-1}\eta}=
\begin{pmatrix}
\frac1{1-l^2}\begin{pmatrix}
1 & l \\
-l & 1-2l^2 \\
\end{pmatrix} && \\
& 1 & \\
& & 1 \\
\end{pmatrix},
\end{equation}
which is indeed real, as it should be.
Therefore our definition of the $\pi$ field yields
\begin{equation}
\phi_0 \rightarrow\frac{\dot\pi-l\pi'}{1-l^2},\qquad
\phi_1 \rightarrow\frac{l\dot\pi+(1-2l^2)\pi'}{1-l^2}.
\end{equation}
If we substitute this into the first term in (\ref{eq:413}) and concentrate on the
second order derivative terms of $\pi$ (forgetting about the
terms with the coefficients differentiated), we have
\begin{eqnarray}
&&\hspace{-2em}\dot\phi_1-l\dot\phi_0 + (2l^2-1)\phi'_0-l\phi'_1 \nn
&=&
\frac{l\ddot\pi+(1-2l^2)\dot\pi'}{1-l^2}
-l\frac{\ddot\pi-l\dot\pi'}{1-l^2}
+ (2l^2-1)
\frac{\dot\pi'-l\pi''}{1-l^2}
-l\frac{l\dot\pi'+(1-2l^2)\pi''}{1-l^2}
=0 \ ! ~~~~~
\end{eqnarray}
Thus we explicitly see that all the second order derivative terms of $\pi$
disappear as was shown generally in the previous section. This is due to the
`gauge invariance' of the $F_{mn}$ term under $\delta\bar\phi_m= \dbar\partial_m\pi$.
This also clearly shows the importance and nontriviality of our definition of
the St\"{u}ckelberg $\pi$-field or decoupling limit in the general curved spacetime.

When $l$ becomes 1, our expression for the quadratic term of the mass term
diverges [see Eq.~\p{eq:413}].
What happens there?

As long as the condition $l^2<1$ is satisfied, the characteristic equation~(\ref{eq:character})
has two roots $\alpha$ and $\alpha^*$, and the matrix $\gbar^{-1}\eta$ is diagonalizable.
But when $l$ becomes as large as 1,
the complex eigenvalues $\alpha$ and $\alpha^*$ become degenerate and
take the value $-1$, and the corresponding eigenvectors $V_1$ and $V_2$ also
degenerate, i.e., $NV_1\propto NV_2$. This implies that
the eigenvectors do not span a complete set so that the matrix $\gbar^{-1}\eta$
is non-diagonalizable. At $l=1$, $\gbar^{-1}\eta$ can be brought
at most into a Jordan standard form:
\begin{equation}
\quad V^{-1} \gbar^{-1}\eta V =
\begin{pmatrix}
-1&1&& \\
& -1&& \\
&&1&\\
&&&1\\
\end{pmatrix}.
\end{equation}
The form of the quadratic kinetic term for the
St\"{u}ckelberg fields, which was derived in the previous section assuming
diagonalizability, diverges in the limit $l\to 1$.

Fortunately the mass term $\frac12 [(K_\mu^\mu)^2-K_\mu^\nu K_\nu^\mu]$ can be calculated
exactly in this example if we retain only the $\partial_\mu\phi_\nu$ terms with
$\mu,\,\nu=0,1$.\footnote{%
In this case the matrix $A=\gbar^{-1}f$ becomes essentially 2-dimensional.
Any $2\times2$ real matrix $A$ can always be written in the form
$A= a_0 1_2+ \vec a \cdot \vec\sigma$
in terms of four parameters $a_\mu$, three real $a_0, \ a_1,\ a_3$ and purely
imaginary $a_2$, together with unit matrix $1_2$ and Pauli matrices
$\vec\sigma=(\sigma_1,\sigma_2,\sigma_3)$.
Using this parametrization and the properties of the Pauli matrices,
one can easily find the square root as
\begin{equation}
\sqrt{A} = b_0 1_2+ \frac1{2b_0}\vec a \cdot \vec\sigma \quad \hbox{with}\quad
2b_0^2 = \frac12 \tr A + \sqrt{\det{A}}
\end{equation}
}
This is fine since we are mainly interested in the time derivatives of the fields.
If we keep only the time derivative terms $\dot\phi_0$ and $\dot\phi_1$,
we find
\begin{equation}
2 - \dot\phi_0 - \sqrt{4(1-l^2) -4 (\dot\phi_0 -l\dot\phi_1) +\dot\phi_0^2 - \dot\phi_1^2}.
\label{eq:jordan}
\end{equation}
If we look at Eq.~\p{eq:jordan} for $l=1$, we see that $\dot\phi_\mu=0$ point
becomes the branch point of the square root so that the expansion itself of the mass term
in powers of the St\"{u}ckelberg fields $\phi$ does not make sense.

The origin of the square root here is of course the square root factor $
\sqrt{g^{-1}f}$ of the dRGT mass term. So even if we do not introduce the
St\"{u}ckelberg fields $\phi$ (i.e., setting $f_{\mu\nu}=\eta_{\mu\nu}$), this
singularity at $g=\gbar$ with $l=1$ is the singularity of the Lagrangian
itself and the metric fluctuation $h_{\mu\nu}$ around the background
$g=\gbar$ does not make sense. This does not allow for any particle
interpretation.

Beyond $|l|=1$ also, the background value inside the square root in
Eq.~\p{eq:jordan} is negative, and again this implies that the
the square root factor
$\sqrt{g^{-1}f}$ in the dRGT mass term comes to have {\em complex} value
at the background $g=\gbar$ so that the dRGT Lagrangian itself becomes
{\em non-hermitian} and no longer gives a well-defined theory.

This is the reason why we have to restrict the shift
vector to $|l|<1$, and in this region our discussions work perfectly
well and there is no ghost in this massive gravity.
This must be the general situation; as long as the dRGT mass term defines
a hermitian Lagrangian, then the matrix ${\gbar^{-1}f}$ is diagonalizable
and the general no ghost proof in the previous section will apply.

In summary,
we have discussed the no-ghost theorem in massive gravity.
We start with the discussion of the simple gravity theory with Fierz-Pauli mass term
and analyze the spectrum in a covariant manner. Naively we have six degrees of freedom
since the general coordinate invariance is broken in the presence of the mass term.
However, we have shown that one of the modes, BD ghost, decouples for the special choice
of the mass term. By introducing the St\"{u}ckelberg fields,
which recover the general coordinate invariance, and using the BRST formalism,
we have then clarified how the various modes in the theory cancels each other,
leaving the correct five degrees of freedom.
The crucial point in this formulation is that there remains no higher (time) derivative
on the St\"{u}ckelberg fields.

We then proceed to the discussion of the nonlinear dRGT massive gravity on arbitrary backgrounds.
Because the complicated nature of the square root form of the mass term, it is rather
cumbersome to identify fluctuations around arbitrary backgrounds, but we were able
to do it by diagonalizing the background. We have then shown that there remains no higher
(time) derivatives on the St\"{u}ckelberg fields, and hence the theory is free from ghost.
In this process, we have identified the correct way to introduce the St\"{u}ckelberg fields
on general backgrounds, and found that the associated decoupling limit is also quite nontrivial,
naively mixing time and space derivatives.
Nonetheless, we have shown that this does not cause trouble with the ghost.
Rather this is necessary in order for the ghost to decouple.
This is further confirmed by an explicit example.

Recently it has been shown that this class of massive gravity can be derived from
the five-dimensional Einstein gravity by deconstruction~\cite{DMT2}.
It would be interesting to extend that approach to supergravity and
study the structure of the theory.

\section*{Acknowledgment}

Part of this work was carried out while the authors were attending the molecule-type
workshop ``Nonlinear massive gravity theory and its observational test'' (ID: YITP-T-12-04).
We thank the organizer (Tetsuya Shiromizu) and participants, in particular Claudia de Rham
and Alndrew Tolley for stimulating discussions.
Thanks are also due to Cedric Deffayet, Fawad Hassan, Kei-ichi Maeda,
Yuho Sakatani, Tomohiko Takahashi for valuable discussions.
We are much indebted especially to Shinji Mukohyama who taught us the
expansion formula for the matrix square root around the diagonal matrix
which was crucial for the present work.
One of the authors (T.K.) is partially supported
by a Grant-in-Aid for Scientific
Research (B) (No.\ 24340049) from the Japan Society for the Promotion of Science (JSPS).
The other author (N.O.) is supported in part by the Grant-in-Aid for
Scientific Research Fund of the JSPS (C) No.~24540290, and (A) No.~22244030.


\begin{thebibliography}{99}

\bibitem{FP}
M.~Fierz and W.~Pauli,
 ``On relativistic wave equations for particles of arbitrary spin in an electromagnetic field,''
  Proc.\ Roy.\ Soc.\ Lond.\ A {\bf 173} (1939) 211.

\bibitem{disc}
  H.~van Dam and M.~J.~G.~Veltman,
  ``Massive and massless Yang-Mills and gravitational fields,''
  Nucl.\ Phys.\ B {\bf 22} (1970) 397;\\
 V.~I.~Zakharov,
  ``Linearized gravitation theory and the graviton mass,''
  JETP Lett.\  {\bf 12} (1970) 312
   [Pisma Zh.\ Eksp.\ Teor.\ Fiz.\  {\bf 12} (1970) 447].

\bibitem{V}
  A.~I.~Vainshtein,
  ``To the problem of nonvanishing gravitation mass,''
  Phys.\ Lett.\ B {\bf 39} (1972) 393.

\bibitem{BD}
  D.~G.~Boulware and S.~Deser,
  ``Can gravitation have a finite range?,''
  Phys.\ Rev.\ D {\bf 6} (1972) 3368.

\bibitem{DG}
  C.~de Rham and G.~Gabadadze,
  ``Generalization of the Fierz-Pauli Action,''
  Phys.\ Rev.\ D {\bf 82} (2010) 044020
  [arXiv:1007.0443 [hep-th]].

\bibitem{DGT1}
  C.~de Rham, G.~Gabadadze and A.~J.~Tolley,
  ``Resummation of Massive Gravity,''
  Phys.\ Rev.\ Lett.\  {\bf 106} (2011) 231101
  [arXiv:1011.1232 [hep-th]].

\bibitem{AGS}
  N.~Arkani-Hamed, H.~Georgi and M.~D.~Schwartz,
  ``Effective field theory for massive gravitons and gravity in theory space,''
  Annals Phys.\  {\bf 305} (2003) 96
  [hep-th/0210184].

\bibitem{HR1}
  S.~F.~Hassan and R.~A.~Rosen,
  ``Resolving the Ghost Problem in non-Linear Massive Gravity,''
  Phys.\ Rev.\ Lett.\  {\bf 108} (2012) 041101
  [arXiv:1106.3344 [hep-th]].

\bibitem{DGT2}
  C.~de Rham, G.~Gabadadze and A.~J.~Tolley,
  ``Ghost free Massive Gravity in the St\"uckelberg language,''
  Phys.\ Lett.\ B {\bf 711} (2012) 190
  [arXiv:1107.3820 [hep-th]].

\bibitem{DGT3}
  C.~de Rham, G.~Gabadadze and A.~J.~Tolley,
  ``Helicity Decomposition of Ghost-free Massive Gravity,''
  JHEP {\bf 1111} (2011) 093
  [arXiv:1108.4521 [hep-th]].

\bibitem{HR2}
  S.~F.~Hassan, R.~A.~Rosen and A.~Schmidt-May,
  ``Ghost-free Massive Gravity with a General Reference Metric,''
  JHEP {\bf 1202} (2012) 026
  [arXiv:1109.3230 [hep-th]].

\bibitem{HR3}
  S.~F.~Hassan and R.~A.~Rosen,
  ``Confirmation of the Secondary Constraint and Absence of Ghost in Massive Gravity
 and Bimetric Gravity,''
  JHEP {\bf 1204} (2012) 123
  [arXiv:1111.2070 [hep-th]].

\bibitem{Mir}
  M.~Mirbabayi,
  ``A Proof Of Ghost Freedom In de Rham-Gabadadze-Tolley Massive Gravity,''
  Phys.\ Rev.\ D {\bf 86} (2012) 084006
  [arXiv:1112.1435 [hep-th]].

\bibitem{HR4}
  S.~F.~Hassan, A.~Schmidt-May and M.~von Strauss,
  ``Proof of Consistency of Nonlinear Massive Gravity in the St\"uckelberg Formulation,''
  Phys.\ Lett.\ B {\bf 715} (2012) 335
  [arXiv:1203.5283 [hep-th]].

\bibitem{HR}
  K.~Hinterbichler and R.~A.~Rosen,
  ``Interacting Spin-2 Fields,''
  JHEP {\bf 1207} (2012) 047
  [arXiv:1203.5783 [hep-th]].

\bibitem{DMZ}
  C.~Deffayet, J.~Mourad and G.~Zahariade,
  ``Covariant constraints in ghost free massive gravity,''
  JCAP {\bf 1301} (2013) 032
  [arXiv:1207.6338 [hep-th]].

\bibitem{H}
  K.~Hinterbichler,
  ``Ghost-Free Derivative Interactions for a Massive Graviton,''
  JHEP {\bf 1310} (2013) 102
  [arXiv:1305.7227 [hep-th]].

\bibitem{KY}
  R.~Kimura and D.~Yamauchi,
  ``Derivative interactions in de Rham-Gabadadze-Tolley massive gravity,''
  Phys.\ Rev.\ D {\bf 88} (2013) 084025
  [arXiv:1308.0523 [gr-qc]].

\bibitem{DMT1}
  C.~de Rham, A.~Matas and A.~J.~Tolley,
  ``New Kinetic Interactions for Massive Gravity?,''
  arXiv:1311.6485 [hep-th].

\bibitem{Ohta}
  N.~Ohta,
  ``A Complete Classification of Higher Derivative Gravity in 3D and Criticality in 4D,''
  Class.\ Quant.\ Grav.\  {\bf 29} (2012) 015002
  [arXiv:1109.4458 [hep-th]].

\bibitem{Hamamoto:2005in}
  S.~Hamamoto,
  ``Possible nonlinear completion of massive gravity,''
  Prog.\ Theor.\ Phys.\  {\bf 114} (2006) 1261
  [hep-th/0505194].

\bibitem{GLM}
  A.~E.~Gumrukcuoglu, C.~Lin and S.~Mukohyama,
  ``Cosmological perturbations of self-accelerating universe in nonlinear massive gravity,''
  JCAP {\bf 1203} (2012) 006
  [arXiv:1111.4107 [hep-th]].

\bibitem{DMT2}
  C.~de Rham, A.~Matas and A.~J.~Tolley,
  ``Deconstructing Dimensions and Massive Gravity,''
  Class.\ Quant.\ Grav.\  {\bf 31} (2014) 025004
  [arXiv:1308.4136 [hep-th]].

\end{thebibliography}
\end{document}